\documentclass[aps,amsmath,twocolumn,amssymb,floatfix,showpacs, footinbib, longbibliography,nofootinbib]{revtex4-2}

\date{\today}
\usepackage{epsfig}
\usepackage{tabularx}
\usepackage{epstopdf}
\usepackage{subfig}
\usepackage{graphicx}
\usepackage{dcolumn}
\usepackage{bm}
\usepackage{amsmath}
\usepackage[colorlinks,linkcolor=blue,hyperindex,CJKbookmarks]{hyperref}
\usepackage{array}
\usepackage{float}
\usepackage{hyperref}
\usepackage{comment}
\usepackage{xcolor}
\usepackage{makecell}
\usepackage{soul}   
\usepackage{todonotes} 
\usepackage{textcomp}

\begin{document}

\title{Emergence of Lissajous trajectories in skyrmion oscillator}

\author{Tamali Mukherjee}
\author{V Satya Narayana Murthy}
\email{satyam@hyderabad.bits-pilani.ac.in}
\affiliation{Department of Physics, Birla Institute of Technology and Science, Pilani, Hyderabad Campus, Jawahar Nagar, Kapra Mandal, Medchal District, Telangana 500078, India}

\date{\today}

\begin{abstract}
Understanding the dynamics of current-driven skyrmion is essential for their practical applications. In this study, we apply an ac pulse (a) in x-- direction, and (b) in both x-- and y-- directions through the free layer of a Co/Pt thin film and investigate the motion of the skyrmion. We show that the skyrmion follows the sinusoidal current pulse and behaves like an oscillator in the range of current amplitude 1 $\times$ 10$^{11}$ A/m$^2$ to 1 $\times$ 10$^{12}$ A/m$^2$ and frequency 5 $\times$ 10$^{8}$ rad/s to 1 $\times$ 10$^{10}$ rad/s. For current pulse of (A$_1$sin$\omega_1$t, A$_2$sin($\omega_2$t+$\phi$), 0), the skyrmion forms Lissajous figures in the x-y plane, same as observed in classical mechanics. The results are compared at T = 0 K and T $>$ 0 K to analyze the effect of temperature. As the skyrmion Hall angle ($\theta_{SkH}$) and stochastic thermal fluctuation ($\bm{F}^{Th}$) are functions of temperature, the skyrmion starts deviating from its path at T = 0 K with increasing temperature and eventually generates somewhat deformed Lissajous figures from ideal.

\end{abstract}

\maketitle

\section{Introduction}
\label{sec_intro}
Since the formulation of skyrmions as topological solitons by Tony Skyrme in the 1960s \cite{SKYRME1962556}, they have been a topic of interest and active research across various physics domains and have recently been discovered in Bose-Einstein condensates \cite{al2001skyrmions, PhysRevA.111.013301}, liquid crystals \cite{PhysRevE.90.042502}, quantum Hall systems \cite{PhysRevB.55.R1934}, and magnetism \cite{heinze2011spontaneous, fert2017magnetic}. The magnetic skyrmion is a topologically stable vortex-like configuration that exhibits particle-like properties \cite{PhysRevB.87.214419, PhysRevB.107.174412} and offer a powerful platform for energy efficient spintronic applications \cite{10.1063/5.0042917, tomasello2014strategy, zhang2015magnetic, luo2018reconfigurable, PhysRevApplied.17.064035, mukherjee2026skyrmion, song2020skyrmion, doi:10.1126/sciadv.abq5652}.
\par Magnetic skyrmions are observed across a wide range of materials \cite{zhang2023magnetic, fert2017magnetic, doi:10.1021/acs.chemrev.0c00297}, and they are stabilized by competing magnetic interactions such as exchange interaction, Dzyaloshinskii-Moriya interaction (DMI) and anisotropy. Primarily known material systems that can hold skyrmions are non-centrosymmetric bulk materials \cite{doi:10.1126/science.1166767}, and thin film multilayers \cite{heinze2011spontaneous}. In contrast to Bloch-type skyrmions found in bulk materials, N\'eel-type skyrmions generated in thin-film multilayers offer higher efficiency for use in modern spintronic devices due to their ultra-small size and response to applied current \cite{https://doi.org/10.1002/adfm.202504100, annurev:/content/journals/10.1146/annurev-conmatphys-031620-110344, 10.1063/5.0223004}. One of the most common and widely studied system in context of N\'eel-type skyrmions over the years is ferromagnet (FM) / heavy metal (HM) multilayer where skyrmions get stability due to the interfacial DMI\cite{sampaio2013nucleation, PhysRevB.90.094410, soumyanarayanan2017tunable, woo2016observation}.

\par Over the past decade, significant progress has been made in advancing magnetic skyrmions toward practical applications \cite{Bo_2022, Zhang_2020} like logic gate \cite{10.1093/nsr/nwac021}, racetrack memory \cite{10.1145/3333336}, neuromorphic computing \cite{D0MH01603A}, skyrmion-based nano-oscillators \cite{Li_2020}. To put the skyrmions into applications, efficient ways of manipulating them must be well understood. The extensive research on utilization of spin-orbit torque (SOT) \cite{10.1063/1.5041793} and spin-transfer torque (STT) \cite{https://doi.org/10.1002/pssr.201700163} have paved a way to realize the skyrmions in motion \cite{PhysRevB.92.174405, LEE201814}. Furthermore, continued advances in material design and control mechanisms have further improved the stability and tunability of skyrmions at a finite temperature, T$>$ 0 K. 

\par As Co/Pt is a well established multilayer system to nucleate skyrmions at T = 0 K as well as at room temperature \cite{yuan2016skyrmion, mukherjee2025role}, in this article, we study the dynamics of a skyrmion driven by an ac pulse. It is highlighted that, although a skyrmion is not a fundamental particle and arises from a collective spin configuration, it displays well-defined particle-like dynamics. This manuscript is structured as follows; section \ref{sec_method} describes the theoretical formulation of skyrmion dynamics obtained from Thiele equation and the computational modeling used, sections \ref{sec_result1}A and \ref{sec_result1}B explain the simulated result at T = 0 K followed by section \ref{sec_result1}C compares the result at a finite temperature. Finally, section \ref{sec_conclu} outlines the notable outcomes in order to understand the dynamics of skyrmions in more detail and put it into potential application in future.

\begin{figure*}[htbp]
\includegraphics[width=0.8\textwidth,angle=0]{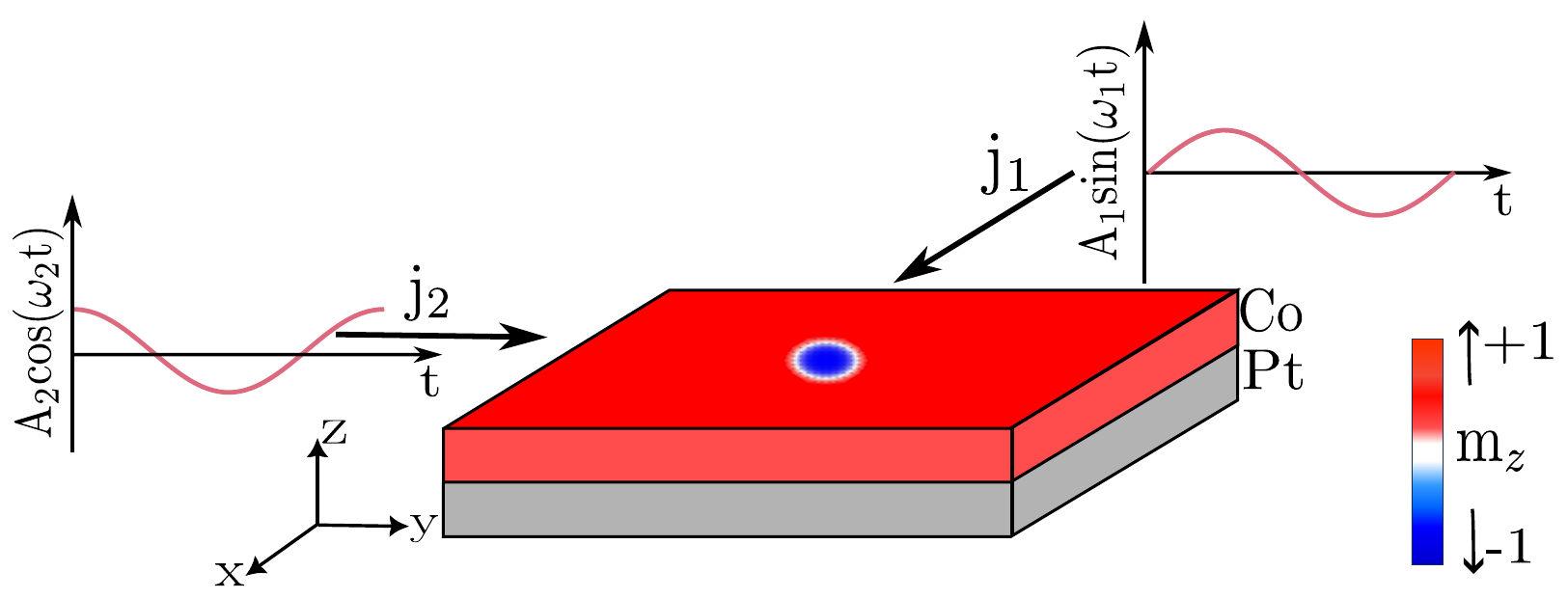}
\caption{ Co/Pt multilayer of (200$\times$200) nm$^2$ is considered where a skyrmion of -z core magnetization is present. An ac pulse of $\bm{j}$ = (j$_1$, j$_2$, 0) = (A$_1$sin($\omega_1$t), A$_2$cos($\omega_2$t), 0) is applied to the nano-structure. }
\label{fig1}
\end{figure*}

\section{Theory and Micromagnetic Modeling}
\label{sec_method}

\par A (200 x 200) nm$^2$ nano-structure of Co/Pt multilayer is considered to study the dynamic behavior of magnetic skyrmions. The 1 nm-thick Co layer is simulated using \textsc{Mumax}$^3$ to solve the Landau-Lifshitz-Gilbert (LLG) equation \cite{10.1098/rsta.2010.0319}. We apply an in-plane current (CIP) through the free layer of Co to induce Zhang Li-type STT, providing the required thrust to drive the skyrmion in the system. The LLG equation along with the Zhang Li type STT is given by,  
\begin{equation}
\frac{d \bm{m}}{d t}=
-\gamma \bm{m}\times\bm{H}_{eff}
-\alpha\bm{m} \times \frac{d \bm{m}}{d t}+\bm{\tau}_{ZL}
\end{equation}
Where, $\gamma$ = gyromagnetic ratio and $\alpha$ = Gilbert damping parameter.
The effective magnetic field (H$_{eff}$) = H$_{exch}$+ H$_{DMI}$ + H$_{demag}$ + H$_{anis}$.

\par The Zhang Li type STT term \cite{PhysRevLett.93.127204} reads as,

\begin{multline}
\bm{\tau}_{\mathrm{ZL}} =
\frac{P \mu_B}{e M_s (1+\beta^2)(1+\alpha^2)}
\Big[
\bm{m} \times \left(\bm{m} \times (\bm{j}\cdot\nabla)\bm{m}\right) \\
+ (\beta - \alpha)\,
\bm{m} \times (\bm{j}\cdot\nabla)\bm{m}
\Big]
\end{multline}

 Here, P is the applied current polarization
 , $\mu_B$ is the Bohr magneton, e is the charge of an electron, M$_s$ is the saturation magnetization, $\beta$ is the non-adiabatic factor, and $\bm{j}$ is the current density applied in the nano-structure. The Thiele equation \cite{PhysRevB.99.104409, mukherjee2026skyrmion} for the CIP scenario is: 
\begin{equation}
\bm{G} \times (\bm{v}^{(s)} - \bm{v}^{(d)})+  \mathcal{D} (\beta\bm{v}^{(s)} - \alpha\bm{v}^{(d)}) + \nabla V = 0
\end{equation}

where, \textbf{G} is the gyromagnetic vector = (0, 0, G) = (0, 0, 4$\pi$Q) and Q is the topological charge of the skyrmion, \textbf{v$^{(s)}$} is the velocity of the conduction electrons due to the spin polarized current, \textbf{v$^{(d)}$} is the drift velocity of the skyrmion, and $\nabla${V} is the force originating from geometric confinement.

Computing from equations (1) and (2), 

\begin{equation}
  \bm{v^{(s)}} = \frac{P\mu_B}{eM_s(1+\beta^2)}\bm{j}  
\end{equation}

Now, as we consider the skyrmion at the centre of the nano-structure and far from edge, we neglect the contribution of V. For the spin polarized current injected in x-direction in the form of a sine wave as, $\bm{j}$ = j$_{x}$ $\hat{e_{x}}$ = Asin($\omega$t)$\hat{e_{x}}$, we note the drift velocity of the skyrmion would be,

\begin{equation}
 \bm{{v}_{x}^{(d)}} = \frac{G^2+\mathcal{D}^2\alpha\beta}{G^2+\alpha^2\mathcal{D}^2} \bm{{v}^{(s)}} 
\end{equation}

\begin{equation}
 \bm{{v}_{y}^{(d)}} = \frac{\mathcal{D}G(\alpha-\beta)}{G^2+\alpha^2\mathcal{D}^2}\bm{{v}^{(s)}} 
 \end{equation}

To make the skyrmion move exactly along the current direction, we have considered the damping parameter and non-adiabatic factor equal ($\alpha$ = $\beta$). Therefore, \textbf{v$^{(d)}_y$}= 0, there is no transverse motion of the skyrmion. As a result, skyrmion Hall angle ($\theta_{SkH}$) is calculated to be,

\begin{equation}
  \theta_{SkH} = tan^{-1}(\frac{v_y^{(d)}}{v_x^{(d)}}) = tan^{-1}[\frac{\mathcal{D}G(\alpha-\beta)}{G^2+\mathcal{D}^2\alpha \beta}] = 0^{\circ}. 
 \end{equation}
 
\par Hence, \textbf{v$^{(d)}_x$} = \textbf{v$^{(s)}$} = v$_0^{(s)}$sin($\omega t$) $\hat{e_{x}}$ where, v$_0^{(s)}$ =$\frac{P\mu_B}{eM_s(1+\beta^2)}$A.

\par Thus, in the presence of an ac pulse, the skyrmion behaves as an oscillator whose displacement varies with time as,
\begin{equation}
x(t)=\int{v_x^{(d)}}dt=-\frac{v_0^{(s)}}{\omega}cos(\omega t)+ c
\end{equation}
To understand the oscillatory behavior of the skyrmion, it is placed at the middle of the square nano-structure, as shown in Fig. \ref{fig1}. Initially, the skyrmion is at rest at t = 0
\begin{equation}
x(t=0)= 0
\end{equation}

Equation (8) becomes

\begin{equation}
x(t)=-\frac{v_0^{(s)}}{\omega}cos(\omega t) + \frac{v_0^{(s)}}{\omega} = \frac{v_0^{(s)}}{\omega}sin(\omega t - \frac{\pi}{2}) + \frac{v_0^{(s)}}{\omega}
\end{equation}
\par Similarly, if an ac pulse is applied in x-- and y-- direction together, as, $\bm{j}$ = (A$_1$sin($\omega_1$t), A$_2$sin($\omega_2$t + $\phi$), 0), the corresponding spin velocities ($\bm{v^{(s)}_1}(t)$ and $\bm{v^{(s)}_2}(t)$) and consequently the drift velocities ( $\bm{v^{(d)}_1}(t)$ and $\bm{v^{(d)}_2}$(t)) in the x-- and y-- directions become respectively,

\begin{equation}
\begin{split}
     v^{(s)}_{1}(t)= v^{(d)}_{1}(t) = \frac{P\mu_B}{eM_s(1+\beta^2)} A_1 sin(\omega_1 t) \\= v_{0,1}^{(s)} sin (\omega_1 t)
\end{split}
\end{equation}
 and,
 \begin{equation}
 \begin{split}
     v^{(s)}_{2}(t)= v^{(d)}_{2}(t) = \frac{P\mu_B}{eM_s(1+\beta^2)} A_2 sin(\omega_2 t + \phi) \\
= v_{0,2}^{(s)} sin (\omega_2t+\phi)
 \end{split}
\end{equation}
Consistently, the skyrmion motion in x-- and y-- direction will be,
\begin{equation}
x(t)=-\frac{v_{0,1}^{(s)}}{\omega_1}cos(\omega_1 t) + \frac{v_{0,1}^{(s)}}{\omega_1}
\end{equation}
and,
\begin{equation}
y(t)= -\frac{v_{0,2}^{(s)}}{\omega_2}cos(\omega_2 t + \phi) + \frac{v_{0,2}^{(s)}}{\omega_2}cos(\phi)
\end{equation}
Consequently, the resultant motion of the skyrmion will be a Lissajous trajectory in the x-y plane.
\par The material parameters used to hold the skyrmion in the sample are, saturation magnetization (M$_s$) = 5.8 $\times$ 10$^5$ A/m, exchange constant (A$_{ex}$) = 1.5 $\times$ 10$^{-11}$ J/m, DMI constant (D$_{int}$) = 3 $\times$ 10$^{-3}$ J/m$^2$ \cite{purnama2015guided, PhysRevApplied.12.064053, 10.1021/acsaelm.2c00122}, anisotropy constant (K$_u$) = 8$\times$ 10$^5$ J/m$^3$ and easy axis is considered in (0, 0, 1) direction. The Gilbert damping parameter is $\alpha$ = 0.1. The Zhang Li-type STT parameters are, P = 0.5, and, $\alpha$ = $\beta$ = 0.1. We perform micromagnetic simulations on the system at T = 0 K using the Dormand-Price (RK45) method and at T $>$ 0 K using the sixth-order Runge-Kutta-Fehlberg (RKF56) solver implemented in \textsc{Mumax}$^3$ \cite{10.1063/1.4899186, 10.1063/1.5003957}.

\section{Results}
\label{sec_result1}

\subsection{Skyrmion oscillation under an ac pulse}

\begin{figure*}[htbp]
\includegraphics[width=1.0\textwidth,angle=0]{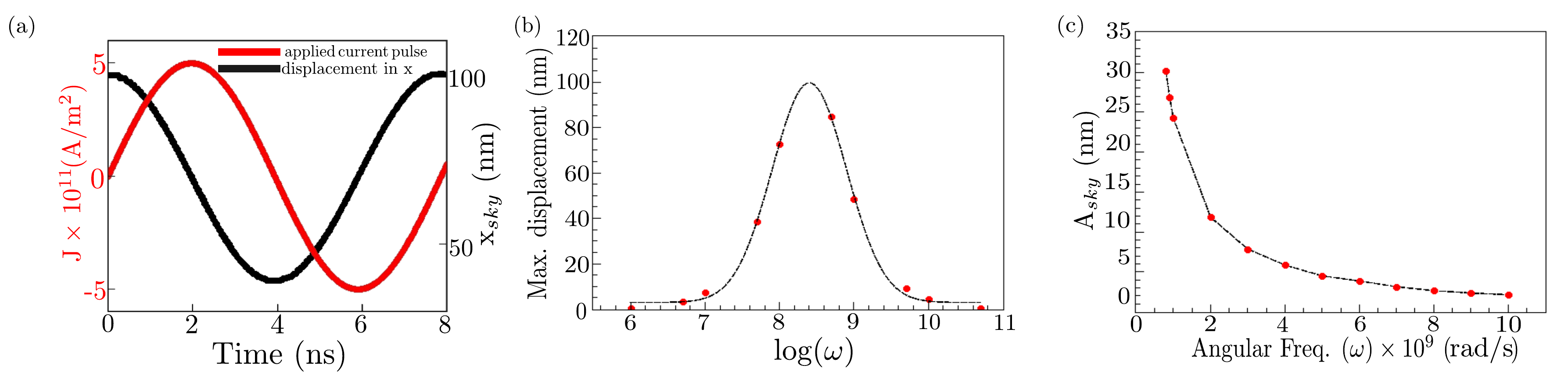}
\caption{ Skyrmion motion under the application of an ac pulse: (a) black line represents the displacement of the skyrmion from the centre of the nano-structure is following a cosine function when $\bm{j}$ = Asin($\omega$t)$\hat{e_x}$ (the red line) where A = 5 $\times$ 10$^{11}$ A/m$^2$ and $\omega$ = 8 $\times$ 10$^8$ rad/s, (b) Distribution of the maximum displacement of the skyrmion with log($\omega$), (c) Variation of A$_{sky}$ with $\omega$. In (b) and (c), red dots denote the simulated data points. The dashed line in (b) corresponds to the Gaussian fit.}
\label{fig2}
\end{figure*}

We keep a skyrmion of radius $\approx$ 8.5 nm and Q = -0.93 exactly at the centre of the nano-structure, (x, y) = (100, 100) and apply an ac pulse of $\bm{j}$ = Asin($\omega$t)$\hat{e_{x}}$ for t = 8 ns. We vary the angular frequency ($\omega$) from 1 $\times$ 10$^6$ rad/s to 1 $\times$ 10$^ {10}$ rad/s, and the amplitude of the pulse (A) in the range of 1 $\times$ 10$^{11}$ A/m$^2$ to 1 $\times$10$^ {12}$ A/m$^2$. The skyrmion begins to oscillate with the current as $\omega$ approaches 10$^8$ rad/s  -- 10$^9$ rad/s within the previously mentioned range of A. Fig. \ref{fig2}(a) depicts the situation where $\bm{j}$ = (5 $\times$ 10$^{11}$)sin((8 $\times$ 10$^8$)t) A/m$^2$ and the skyrmion's trajectory can be described as, 
\begin{equation}
x_{sky}(t) = A_{sky}cos(w_{sky}t)
\end{equation}

Here, x$_{sky}$ denotes the absolute position of the skyrmion in the sample. We observe the skyrmion to move periodically in the x -- direction where A$_{sky}$ is measured as (x$_{max}$ - x$_{min}$)/2 and $\omega_{sky}$ = 2$\pi$/T$_{sky}$. We observe, at $\bm{j}$ = (5 $\times$ 10$^{11}$)sin((8 $\times$ 10$^8$)t) A/m$^2$, the skyrmion oscillates with A$_{sky}$ = 30.48 nm in x -- direction and $\omega_{sky}$ = 0.795 $\times$ 10$^9$ rad/s (Fig. \ref{fig2}a). Fig. \ref{fig2}b shows that the maximum displacement of the skyrmion from the centre of the nano-structure.  The displacement first increases and reaches a maximum and then eventually falls with the increase of $\omega$ of the driving current. The variation in displacement shows a normal distribution with respect to log($\omega$). Furthermore, fig. \ref{fig2}c describes how A$_{sky}$ decreases with the increase in $\omega$.
\par
\begin{figure*}[htbp]
\includegraphics[width= 0.9\textwidth,angle=0]{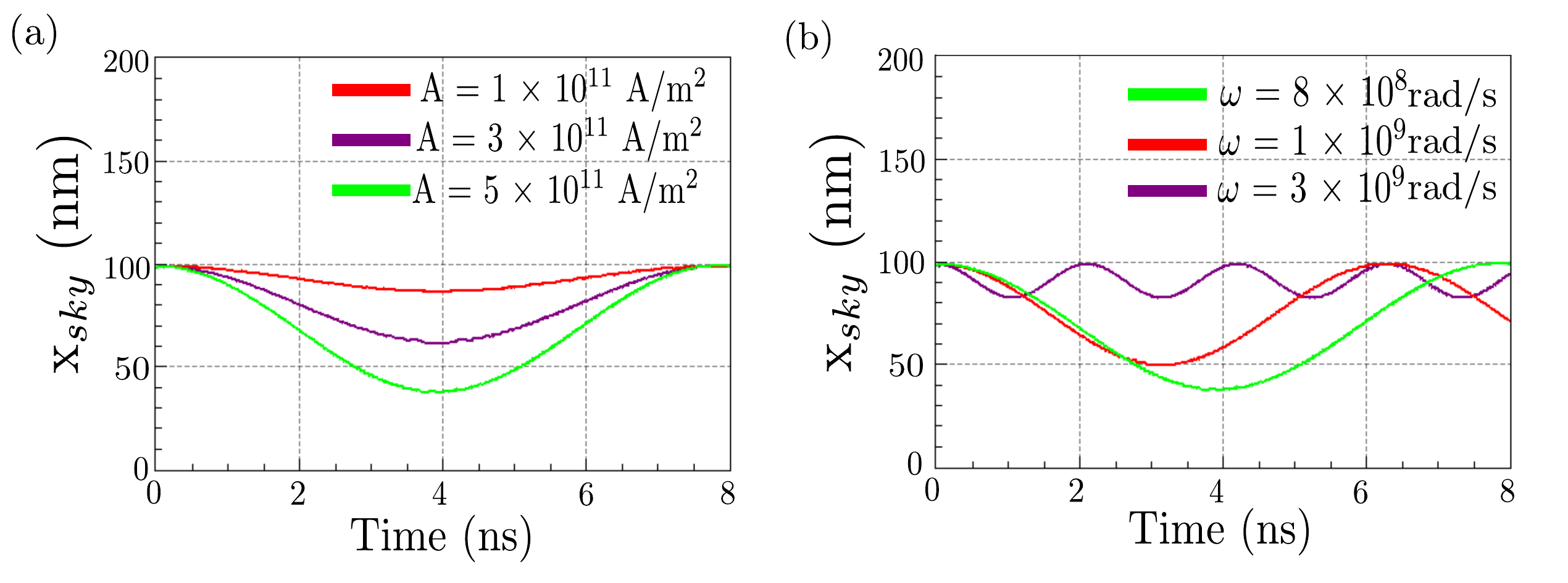}
\caption{ Skyrmion oscillation under the application of $\bm{j}$ =(Asin($\omega t$), 0, 0) when, (a) A = 1 $\times$ 10$^{11}$ A/m$^2$, 3 $\times$ 10$^{11}$ A/m$^2$ and 5 $\times$ 10$^{11}$ A/m$^2$ for a constant $\omega$ = 8 $\times$ 10$^{8}$ rad/s, and in (b) $\omega$ = 8 $\times$ 10$^{8}$ rad/s, 1 $\times$ 10$^{9}$ rad/s and 3 $\times$ 10$^{9}$ rad/s for a constant A = 5 $\times$ 10$^{11}$ A/m$^2$.}
\label{fig3}
\end{figure*}

 Moreover, our analysis indicates that the oscillation of the skyrmion depends on the combined effects of both A and $\omega$ of the driving current $\bm{j}$. Fig. \ref{fig3}a shows the variation in the skyrmion displacement for different applied A at a constant $\omega$ of 8$\times$10$^8$ rad/s. It demonstrates that, as A increases, the skyrmion's speed increases. From eq. (15), the speed of the skyrmion is given by,
\begin{equation}
v_{sky}(t) = -A_{sky}w_{sky}sin(w_{sky}t) 
\end{equation}

The r.m.s values of speed of the skyrmion for A = 1 $\times$ 10$^{11}$ A/m$^2$, 3 $\times$ 10$^{11}$ A/m$^2$, and 5 $\times$ 10$^{11}$ A/m$^2$ increases as 3.51 m/s, 10.30 m/s and 16.92 m/s respectively.

\par Besides, at a constant applied A = 5 $\times$10$^{11}$ A/m$^2$, as we increase $\omega$, consequently $\omega_{sky}$ also increases (Fig. \ref{fig3}b). $\omega_{sky}$ is calculated to be 7.95 $\times$ 10$^8$ rad/s for $\omega$ = 8 $\times$ 10$^8$ rad/s, 0.96 $\times$ 10$^9$ rad/s for $\omega$ = 1 $\times$ 10$^9$ rad/s, and 2.85 $\times$ 10$^9$ rad/s for $\omega$ = 3 $\times$ 10$^9$ rad/s. As the skyrmion oscillator is almost following the external drive frequency, this range of $\omega$$_{sky}$ can be indicated as the region of maximum response. At this range of A and $\omega$ applied, from the Figs. \ref{fig2}b and \ref{fig3}b, the frequency at which the displacement is maximum \cite{10.1063/5.0004649}, we denote that frequency as the maximum response frequency ($\omega$$_{sky}^{max. response}$) = 7.95 $\times$ 10$^8$ rad/s. However, this $\omega$$_{sky}^{max. response}$ of oscillation is observed to vary slightly with A and $\omega$ of the external drive, suggesting the non-linear behavior of the system (see supplementary fig. 1).
\par To understand the effect of size of the skyrmion on its oscillatory behavior we vary the DMI value from 3 $\times$ 10$^{-3}$ J/m$^2$ to 3.4 $\times$ 10$^{-3}$ J/m$^2$ by keeping all the other parameters fixed. Due to this change in D$_{int}$, we obtain a skyrmion of radius 14.6 nm and topological charge = - 0.91. This skyrmion is also subjected to motion under the application of ac drive $\bm{j}$ = Asin($\omega$t)$\hat{e_{x}}$ and the results are compared with the previous scenario (for D$_{int}$ = 3 $\times$ 10$^{-3}$ J/m$^2$). The bigger skyrmion shows a slight displacement in the transverse direction along with the oscillatory motion in applied current direction (discussed in the supplementary fig. 2). However, the skyrmion motion along the direction of the applied ac pulse remains the same.

\begin{figure*}[htbp]
\includegraphics[width=1.0\textwidth,angle=0]{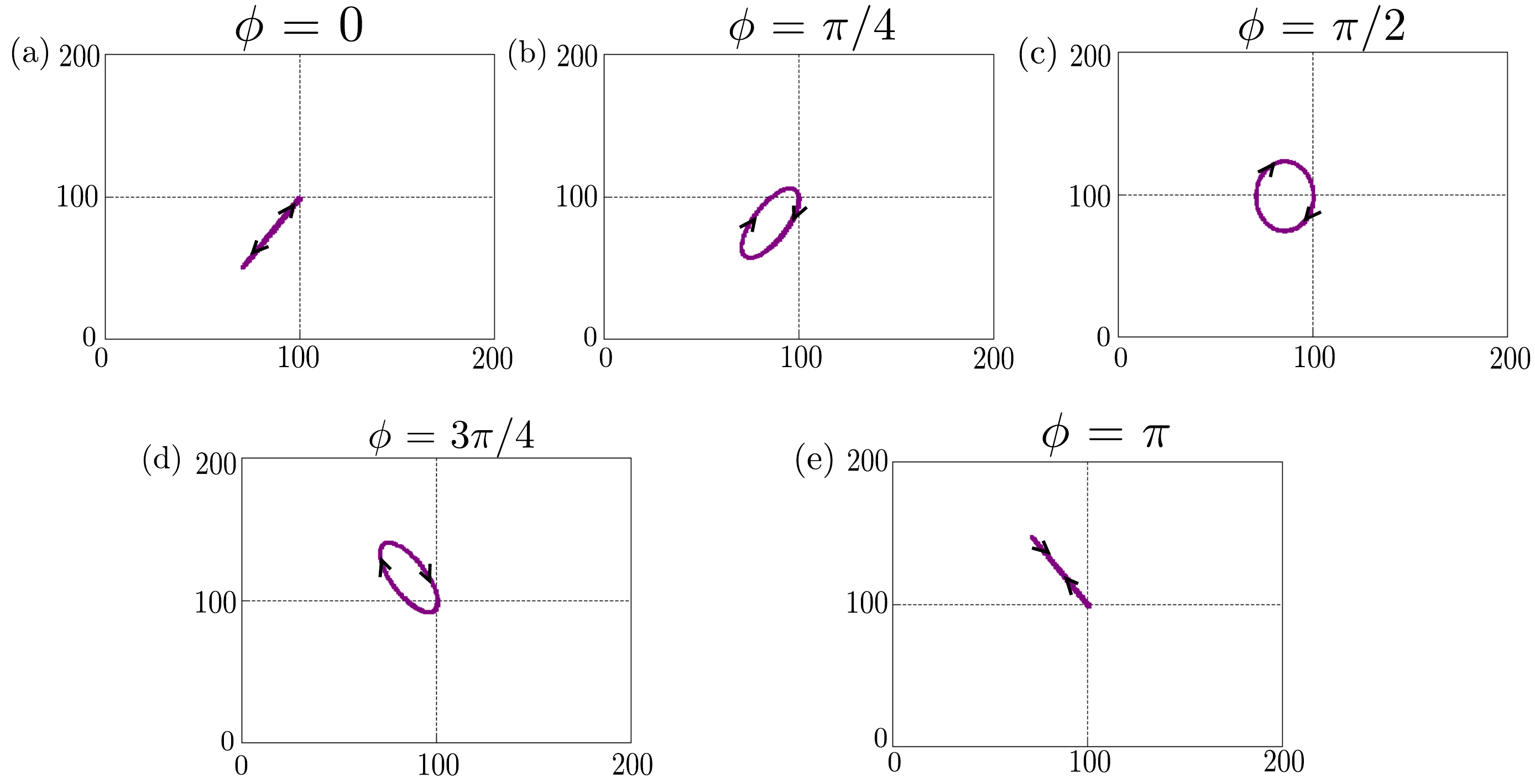}
\caption{ Skyrmion is subjected to motion under (A$_1$sin($\omega t$), A$_2$sin($\omega t$ + $\phi$) , 0) in (200 $\times$ 200) nm$^2$ nano-structure where, A$_1$ = 3 $\times$ 10$^{11}$ A/m$^2$, A$_2$ = 5 $\times$ 10$^{11}$ A/m$^2$, and $\omega$ = 1 $\times$ 10$^{9}$ rad/s. The Lissajous figures obtained are for (a) $\phi$ = 0, (b) $\phi$ = $\pi / 4$, (c) $\phi$ = $\pi /2$, (d) $\phi$ = 3$\pi /4$, and (e) $\phi$ = $\pi$. }
\label{fig4}
\end{figure*}

\begin{table*}
\centering
\renewcommand{\arraystretch}{1.2}

\begin{tabular*}{\textwidth}{@{\extracolsep{\fill}} c c c }
\hline\hline
Phase ($\phi$) & Shape in x-y plane & Meaningful parameters\\
\hline\hline

0 & Straight line & slope = $\tan^{-1}(\frac{A_{2,sky}}{A_{1,sky}}) = 59.21^{\circ}$ \\

$\pi/4$ & Ellipse & tilt = $\frac{1}{2}\tan^{-1}(\frac{2A_{1,sky}A_{2,sky}cos(\pi/4)}{A_{1,sky}^2-A_{2,sky}^2}) = -26.28^{\circ}$ \\

$\pi/2$ & Ellipse & tilt = $\frac{1}{2}\tan^{-1}(\frac{2A_{1,sky}A_{2,sky}cos(\pi/2)}{A_{1,sky}^2-A_{2,sky}^2}) = 0^{\circ}$ \\

$3\pi/4$ & Ellipse & tilt = $\frac{1}{2}\tan^{-1}(\frac{2A_{1,sky}A_{2,sky}cos(3\pi/4)}{A_{1,sky}^2-A_{2,sky}^2}) = 26.28^{\circ}$ \\

$\pi$ & Straight line & slope = $\tan^{-1}(-\frac{A_{2,sky}}{A_{1,sky}}) = -59.21^{\circ}$ \\
\hline\hline
\end{tabular*}

\caption{Parameters obtained from the Lissajous figures}
\label{tab1}
\end{table*}

\subsection{Skyrmion motion under simultaneous x-- and y-- direction ac pulse}

Instead of applying the ac pulse in x-- direction only we further apply the current in both x-- and y-- directions simultaneously to see how the skyrmion behaves.
\par As expected from the perspective of a mechanical particle, Fig. \ref{fig4} ensures that a skyrmion also traces out Lissajous trajectory subjected to the application of two perpendicular ac pulses through the free layer of Co, such as $\bm{j}$ =  (A$_1$sin($\omega$t), A$_2$sin($\omega$t + $\phi$), 0). We vary the $\phi$ from 0 to $\pi$ in $\pi/4$ steps and obtain the Figs. \ref{fig4}a - \ref{fig4}e.
\par As discussed earlier in section \ref{sec_result1}A, for the current pulses mentioned above, the motion of the skyrmion in x-- and y-- directions will respectively be,
\begin{equation}
x_{sky}(t) = A_{1,sky}cos(w_{sky}t)
\end{equation}

and,

\begin{equation}
y_{sky}(t) = A_{2,sky}cos(w_{sky}t + \phi)
\end{equation}

For $\omega$ = 1$\times$ 10$^9$ rad/s, A$_1$ = 3 $\times$ 10$^{11}$ A/m$^2$ and A$_2$ = 5 $\times$ 10$^{11}$ A/m$^2$, A$_{1,sky}$ is observed to be 14.68 nm and A$_{2,sky}$ = 24.64 nm. The skyrmion is tracing out straight lines for $\phi$ = 0 and $\pi$ (Figs. \ref{fig4}a and \ref{fig4}e) and ellipses for $\phi$ = $\pi/4$, $\pi/2$ and $3\pi/4$ in x-y plane (Figs. \ref{fig4}b - \ref{fig4}d). The useful parameters for all these five trajectories in Fig. \ref{fig4} are outlined in table \ref{tab1}.
\par
\begin{figure*}[htbp]
    \includegraphics[width=0.95\textwidth,angle=0]{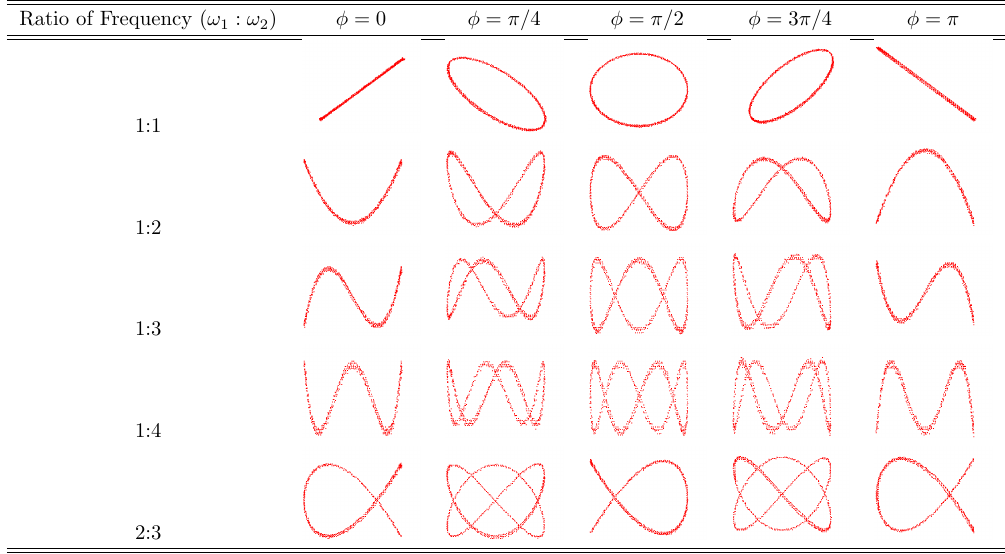}
\caption{Skyrmion Lissajous figures with various phase differences ($\phi$) for various frequency ratios ($\omega_1$:$\omega_2$). The applied current pulse is $\bm{j}$ = (Asin($\omega_1$t), Asin($\omega_2$t + $\phi$), 0) where A = 1 $\times$ 10$^{12}$ A/m$^2$. The snapshots are zoomed-in views of (x$_{sky}$, y$_{sky}$) trajectories, as depicted in Fig. \ref{fig4}.}
\label{tab2}
\end{figure*}

 Now, we apply the current pulse as, $\bm{j}$ = (Asin($\omega_1$t), Asin($\omega_2$t + $\phi$), 0) where $\omega$ we vary in ratio of $\omega_1$ : $\omega_2$ = 1:1, 1:2, 1:3, 1:4 and 2:3 and as a result the skyrmion traces the path in x-y plane as shown in Fig. \ref{tab2}.
Here, A = 1 $\times$ 10$^{12}$ A/m$^2$ and the pulse duration is 8 ns. To obtain these figures, we apply $\omega_1$ = 2 $\times$ 10$^9$ rad/s. The vertical and horizontal lobes present in the figures in Fig. \ref{tab2} perfectly characterize the frequency ratio, and the geometry confirms the phase difference of the ac pulses applied in x-- and y-- directions. These results illustrate the similar behavior of a mechanical particle under an applied ac pulse and elucidate the classical dynamical  characteristics of a magnetic skyrmion.

\par
\begin{figure*}[htbp]
    \includegraphics[width=1.0\textwidth,angle=0]{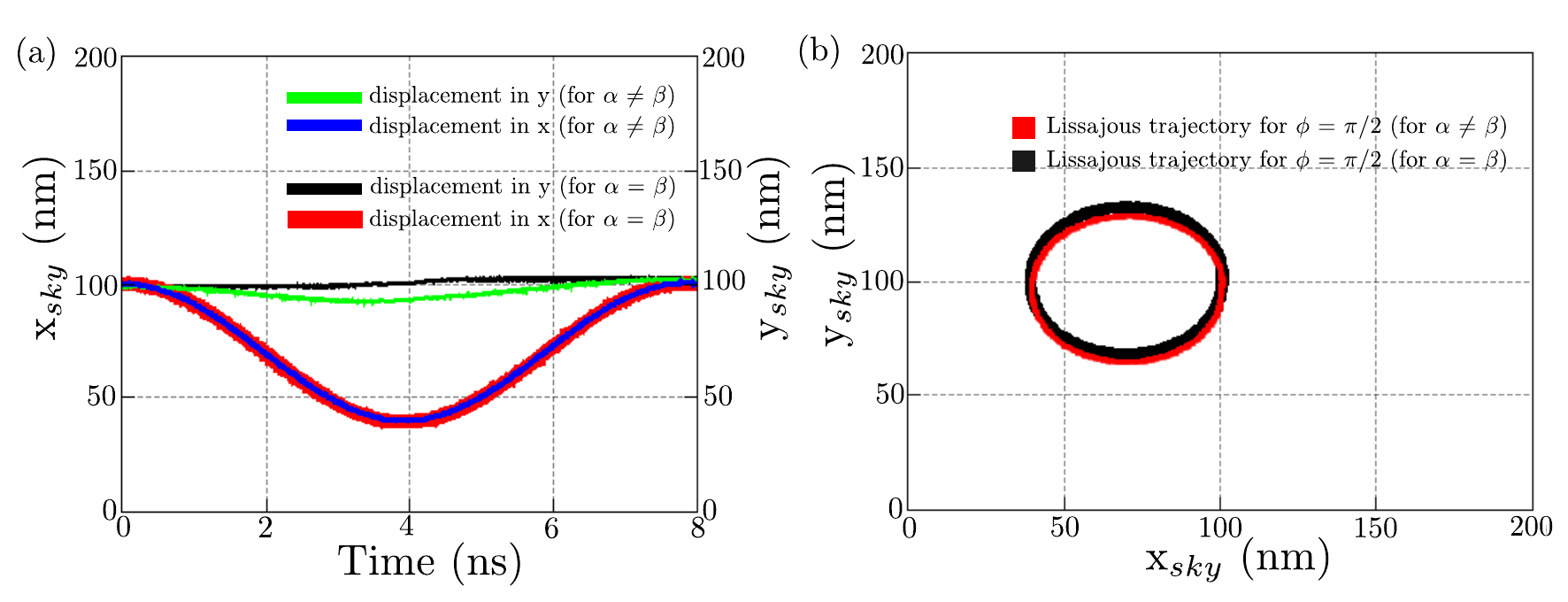}
\caption{Skyrmion motion for $\alpha$ = $\beta$ ($\alpha$ = $\beta$ = 0.1) and $\alpha \neq \beta$ ($\alpha$ = 0.1, $\beta$ = 0.2): (a) displacement in x-- and y-- direction as a function of time under the application of $\bm{j}$= Asin($\omega t$) $\hat{e_{x}}$ for 8 ns, (b) circular Lissajous trajectory obtained for the pulse $\bm{j}$= (Asin($\omega t$), Asin($\omega t+\pi/2$), 0) applied. For both the cases, $\alpha = \beta$ and $\alpha \neq \beta$, A = 5 $\times$ 10$^{11}$ A/m$^2$ and $\omega$ = 8 $\times$ 10$^8$ rad/s.}
\label{fig5}
\end{figure*}

 Additionally, we analyze the skyrmion motion under the condition of $\alpha \neq \beta$ (for $\alpha = 0.1$ and $\beta=0.2$), at first, by injecting $\bm{j}$ = Asin$\omega$t $\hat{e_{x}}$ (Fig. \ref{fig5}a) and then $\bm{j}$ = (Asin($\omega$t), Asin($\omega t+\pi/2$), 0) (Fig.\ref{fig5}b). We find that Lissajous figures obtained for $\phi$ = 0, $\pi/4$, $\pi/2$, $3\pi/4$ and $\pi$ are in well agreement with the trajectories obtained at $\alpha = \beta$. As the Hall angle induced in x$_{sky}$ and y$_{sky}$ due to $\alpha \neq \beta$ condition is the same, the superposition of x-- and y-- direction motions retains the shape of Lissajous figures perfectly.

\par
\begin{figure*}[htbp]
    \includegraphics[width=1.0\textwidth,angle=0]{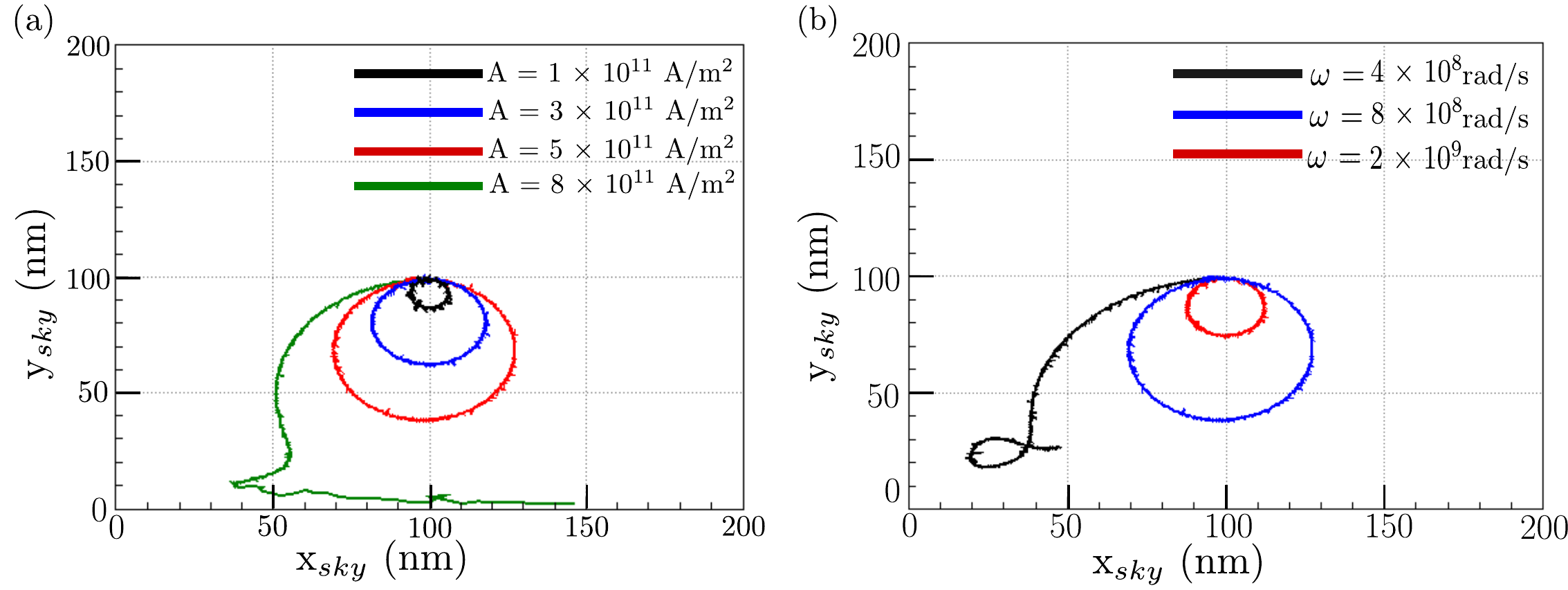}
\caption{Skyrmion Lissajous figures obtained for (a) varying amplitude (A) of ac pulse in the range of 1 $\times$ 10$^{11}$ A/m$^2$ -- 8 $\times$ 10$^{11}$ A/m$^2$ (b) varying angular frequency ($\omega$) of ac pulse in the range of 4 $\times$ 10$^8$ rad/s -- 2 $\times$ 10$^9$ rad/s. The applied ac pulse is $\bm{j}$ = (Acos($\omega t$), Asin($\omega t$), 0) }
\label{fig6}
\end{figure*}

 We notice that for a fixed angular frequency ($\omega$), increasing the amplitude of the driving ac pulse increases the skyrmion's speed and, consequently, its displacement from the center of the nano-structure (Fig. \ref{fig3}a). Similarly, when we apply a current pulse of (Acos($\omega t$), Asin($\omega t$), 0), the diameter of the circular Lissajous trajectory of the skyrmion becomes bigger with increasing driving current amplitude from 1 $\times$ 10$^{11}$ A/m$^2$ to 8 $\times$ 10$^{11}$ A/m$^2$ for a fixed $\omega$ = 8 $\times$ 10$^{8}$ rad/s, as explained in Fig. \ref{fig6}a. On the other hand, when we increase the angular frequency ($\omega$) while keeping the amplitude (A) fixed, we observe the Lissajous circular trajectories to be of smaller size, as expected from the earlier description of skyrmion oscillation (Fig. \ref{fig3}b). Fig. \ref{fig6}(b) demonstrates the condition where $\omega$ is increased from 4 $\times$ 10$^{8}$ rad/s to 2 $\times$ 10$^{9}$ rad/s for A = 5 $\times$ 10$^{11}$ A/m$^2$. Moreover, Fig. \ref{fig6}a and \ref{fig6}b indicate the effect of confinement potential (V) on the skyrmion Lissajous figures. When the skyrmion goes near to the edge of the nano-structure due to sufficiently strong driving ac pulse, the approximation of $\nabla{V} \approx$ 0 does not hold, and the skyrmion cannot follow the Lissajous path due to the edge-induced confinement.

\par To study the effect of the thickness of the free layer Co on the skyrmion Lissajous trajectory we perform the simulations on (200 $\times$ 200 $\times$ 5) nm$^3$ nano-structure. Upon relaxing the system with the set of material parameter values mentioned above, we obtain a skyrmion of radius $\approx$ 10 nm. We apply an ac pulse of (a) $\bm{j}$ = Asin($\omega$t)$\hat{e_{x}}$ and (b) $\bm{j}$ = (Acos($\omega$t), Asin($\omega$t), 0) to the sample through the free layer and outline the dynamics of the skyrmion (supplementary fig. 3). A slight transverse displacement arises for both the applied ac drive in x-- and y-- direction, indicating no significant change in the resultant Lissajous trajectory (supplementary fig. 3a -- 3b).

\subsection{Skyrmion dynamics at finite temperature (T $>$ 0 K)}
To analyze the dynamics of a skyrmion subjected to an ac pulse at a higher temperature, we further apply $\bm{j}$ = (Asin($\omega$t), 0, 0) through the free layer of Co for 8 ns. The modified Thiele equation at a finite temperature T is given by \cite{PhysRevLett.127.047203,JIANG2022169786},

\begin{equation}
\bm{G} \times (\bm{v}^{(s)} - \bm{v}^{(d)})+  \mathcal{D} \beta\bm{v}^{(s)} - (\alpha\mathcal{D} +\eta T)\bm{v}^{(d)} + \nabla V + \bm{F}^{Th} = 0
\end{equation}

\begin{figure*}[htbp]
\includegraphics[width=1.0\textwidth,angle=0]{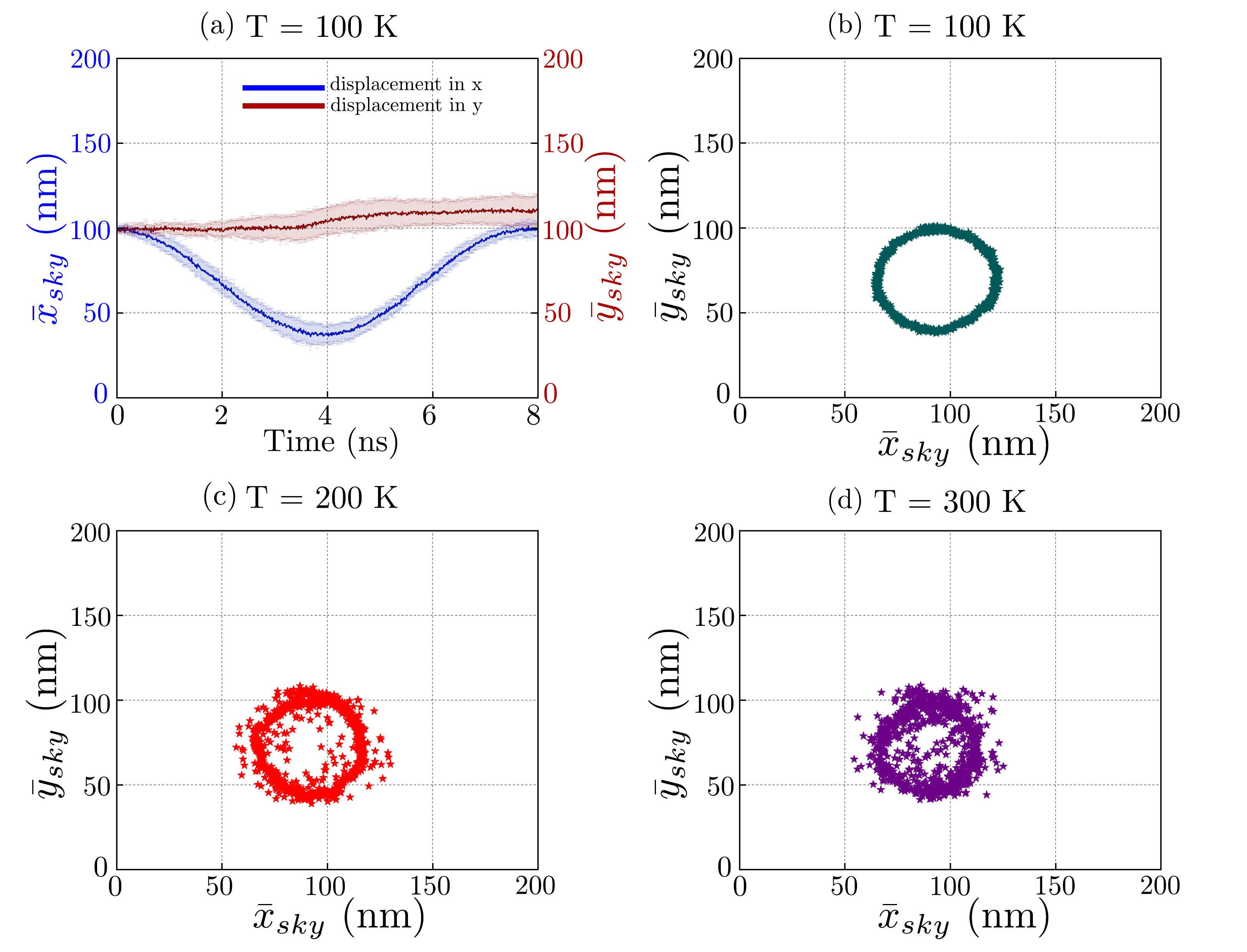}
\caption{ Skyrmion motion at finite temperature: (a) the displacement of the skyrmion for $\bm{j}$ = Asin($\omega$t)$\hat{e_x}$ when A = 5 $\times$ 10$^{11}$ A/m$^2$ and $\omega$ = 8 $\times$ 10$^8$ rad/s, at T = 100 K. (b)--(d) Circular Lissajous trajectory of skyrmion for $\bm{j}$ = (Acos($\omega$t), Asin($\omega$t), 0) where A = 5 $\times$ 10$^{11}$ A/m$^2$ and $\omega$ = 8 $\times$ 10$^8$ rad/s, at T = 0 K, 100 K, 200 K and 300 K. The solid curves denote the ensemble-averaged trajectory. The faded lines of the same color in (a) represent the standard deviation($\pm \sigma$) about the mean trajectory in both the x-- and y--directions for skyrmion oscillation at T = 100 K. }
\label{fig7}
\end{figure*}

Comparing with eq.(3) the additional terms that arise due to high temperature are ($\alpha \mathcal{D}$ + $\eta$T) instead of only $\alpha \mathcal{D}$ and $\bm{F}^{Th}$ that describes the stochastic thermal fluctuation present at T $>$ 0 K. To explain the skyrmion dynamics at an elevated temperature, we must consider the contribution of $\eta T$ in the effective friction of skyrmion in the Thiele eq. As discussed in \cite{PhysRevLett.127.047203}, $\eta T$ corresponds to the magnon-induced friction present in the system that refers to the coupling of skyrmion motion to the magnons at a finite temperature. $\eta$ = 5.05 $\times$ 10$^{-17}$ kg/(sk) and it is independent of T and $\alpha$. Besides, the thermal force follows, $< \bm{F}_{\mu}^{Th} >$ = 0 and $< \bm{F}_{\mu}^{Th}(t)\bm{F}_{\nu}^{Th}(t^\prime) >$ = 2k$_B$T($\alpha\mathcal{D}$ + $\eta T$)$\delta_{\mu\nu}\delta(t-t^\prime)$.
\par Replacing $\alpha\mathcal{D}$ with $(\alpha\mathcal{D} + \eta T)$ in eq.(7) the Hall angle $(\theta_{SkH}$) becomes,

\begin{equation}
  \theta_{SkH} = tan^{-1}(\frac{<v_y^{(d)}>}{<v_x^{(d)}>}) = tan^{-1}[G\frac{(\alpha\mathcal{D} + \eta T)-\beta\mathcal{D}}{G^2+(\alpha\mathcal{D}+ \eta T)\beta\mathcal{D}}]  
 \end{equation}

Thus, even with the condition of  $\alpha$ = $\beta$, $\theta_{SkH}$ $\neq$ 0 and it depends on T. Consequently, the skyrmion has a finite displacement in the y--direction too. To study the effect of temperature on skyrmion motion, we performed ten independent simulations using an identical set of material parameters with different thermal noise realizations, representing ten statistical ensembles and plot the mean position of the skyrmion ($\bar{x}_{sky}$, $\bar{y}_{sky}$) at each time interval of t = 10$^{-12}$ s for T = 100 K, 200 K and 300 K. Fig. \ref{fig7}a illustrates the displacement in both x-- and y-- direction when A = 5$\times$ 10$^{11}$ A/m$^2$ and $\omega$ = 8$\times$ 10$^8$ rad/s at T = 100 K. Here, $\bar{A}_{sky}$(x-- direction, T = 100 K) $\approx$ 29.50 nm and $\bar{A}_{sky}$(y-- direction, T = 100 K) $\approx$ 5 nm. Furthermore, we show the error in absolute skyrmion position (x$_{sky}$, y$_{sky}$) at each time step t, represented by $\pm \sigma$ in Fig. \ref{fig7}a where $\sigma$ denotes the standard deviation.

\par Furthermore, if we apply the ac pulse in both x-- and y-- direction as, $\bm{j}$ = (Acos($\omega$t), Asin($\omega$t), 0), the skyrmion traces out circle in the x-y plane, as expected. However, due to the presence of non-zero $\theta_{SkH}$ and $\bm{F}^{Th}$, the skyrmion trajectories will be distorted from the perfect circular shape.  Figs. \ref{fig7}b, \ref{fig7}c and \ref{fig7}d represents the mean circular trajectories of the skyrmion for T = 100 K, 200 K and 300 K where A = 5$\times$ 10$^{11}$ A/m$^2$ and $\omega$ = 8$\times$ 10$^8$ rad/s. Hence, from the Fig. \ref{fig7}b -- \ref{fig7}d, it is clear that the more the temperature is, the fluctuation is more and more distorted the circle becomes. We demonstrate the error in  (x$_{sky}$, y$_{sky}$) for T = 100 K and 200 K in supplementary fig. 4. Moreover, we describe the fluctuation of the radius of the skyrmion (r$_{sky}$) and topological charge during the entire simulation time for T = 100 K, 200 K and 300 K in the supplementary fig. 5.

\section{Conclusion and outlook}
\label{sec_conclu}
In summary, we outline the dynamics of a magnetic skyrmion subjected to a sinusoidal ac pulse applied through the free layer and characterize its motion at T = 0 K, 100 K, 200 K, and 300 K.  The Thiele equation at T = 0 K and T $>$ 0 K reasonably describes the skyrmion's motion under an ac pulse applied along one direction and under the superposition of two perpendicular ac drives that differ in phase and frequency. The skyrmion is observed to demonstrate the same nature as a mechanical particle under an ac drive. From the results, we infer that the skyrmion exhibits oscillatory behavior and follows a frequency close to that of the driving ac pulse while confined within the mentioned sample geometry, provided the applied ac pulse is of A = 10$^{11}$ A/m$^2$ - 10$^{12}$ A/m$^2$ and $\omega$ = 10$^8$ rad/s - 10$^9$ rad/s. Moreover, the skyrmion traces out Lissajous figures in the x-y plane in response to the superposition of x-- and y-- direction current pulses. These Lissajous patterns formed by the skyrmion carry information about the amplitude, frequency, and phase of the applied current pulse, as expected for a classical particle. Hence, these Lissajous trajectories can be used in future to develop spintronic devices such as a skyrmion frequency comparator (to detect and compare two perpendicular signals), a skyrmion microwave analyzer (to measure high-frequency radio signals), and a phase detector. We examine the skyrmion motion by implying both the conditions $\alpha = \beta$ and $\alpha \ne \beta$ for a fixed driving ac pulse. Future work may explore the dynamics in systems exhibiting negligible skyrmion Hall effect, such as synthetic antiferromagnetic skyrmions.  However, due to the non-zero Hall angle induced at finite temperature (T $>$ 0 K) and thermal fluctuation, the Lissajous figures start to deform from the perfect shapes obtained at T = 0 K.

\section*{Acknowledgment}
\vspace{-1.05 em}
The authors acknowledge the support of BITS Pilani, Hyderabad campus in providing access to the Sharanga High-Performance Computational facility.

\bibliography{lissajous}{}

\clearpage

\onecolumngrid

\section*{Supplementary}

\twocolumngrid

\subsection{Maximum Response frequency of the skyrmion oscillator}

\par
\begin{figure}[htbp]
\includegraphics[width=0.5\textwidth,angle=0]{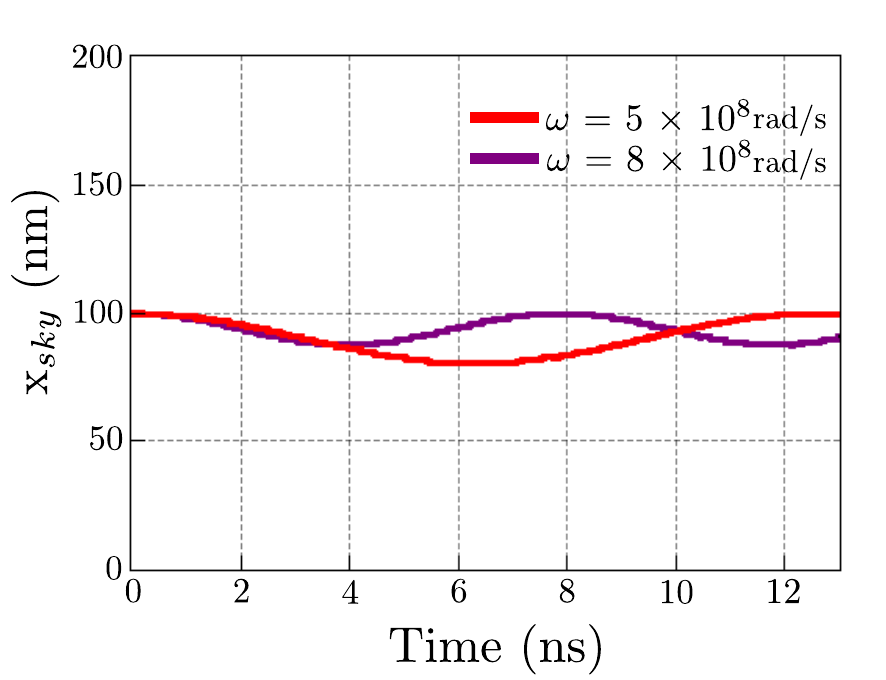}
\caption{{Skyrmion oscillation for $\bm{j}$ = (Asin($\omega t$), 0, 0) where A = 1 $\times$ 10$^{11}$ A/m$^2$ and $\omega$ increased from 5 $\times$ 10$^8$ rad/s to 8 $\times$ 10$^8$ rad/s.}}
\label{sfig1}
\end{figure}

 We observe the skyrmion to start oscillating in 200 $\times$ 200 nm$^2$ nano-structure at $\bm{j}$ = (Asin($\omega t$), 0, 0) when A = 1 $\times$ 10$^{11}$ A/m$^2$. At this magnitude of current, we observe the skyrmion to oscillate with maximum amplitude where $\omega$ = 5 $\times$ 10$^8$ rad/s as shown in Fig. \ref{sfig1}. Here, A$_{sky}$ varies from 9.75 nm to 5.64 nm for $\omega$ = 5 $\times$ 10$^8$ rad/s and $\omega$ = 8 $\times$ 10$^8$ rad/s respectively. At A = 1 $\times$ 10$^{11}$ A/m$^2$, $\omega_{sky}^{max. response}$ $\approx$ 5 $\times$ 10$^8$ rad/s. 

\par We find that the oscillation of the skyrmion confined within the sample's geometry is influenced by parameters such as A and $\omega$ of the applied current pulse. Moreover, Fig. \ref{sfig1} confirms that A$_{sky}$, and consequently $\omega_{sky}^{max. response}$ depends on A, demonstrating the nonlinearity of the skyrmion oscillator.
In summary, we notice that decreasing A from 5 $\times$ 10$^{11}$ A/m$^2$ to 1 $\times$ 10$^{11}$ A/m$^2$, results in a reduction of $\omega_{sky}^{max. response}$ from 8 $\times$ 10$^{8}$ rad/s to 5 $\times$ 10$^{8}$ rad/s.

\begin{figure}[htbp]
\includegraphics[width=0.5\textwidth,angle=0]{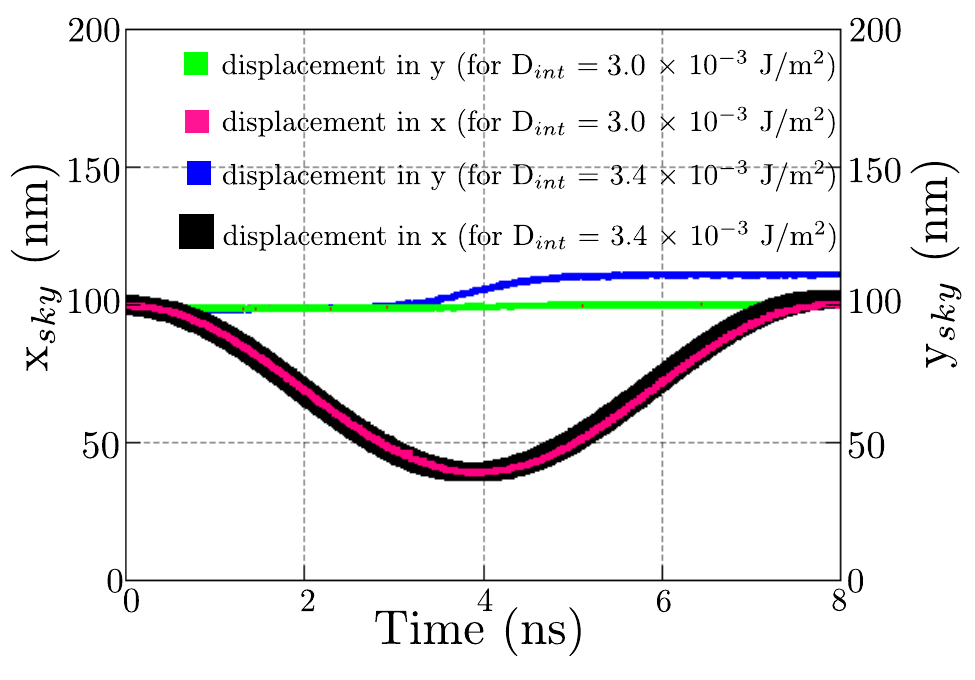}
\caption{{The skyrmions of radii 8.5 nm (for D$_{int}$ = 3.0 $\times$ 10$^{-3}$ J/m$^3$ and Q = -0.93) and 14.6 nm (for D$_{int}$ = 3.4 $\times$ 10$^{-3}$ J/m$^3$ and Q = -0.91) respectively are subjected to motion under $\bm{j}$ = (Asin($\omega t$), 0, 0) where A = 5 $\times$ 10$^{11}$ A/m$^2$ and $\omega$ = 8 $\times$ 10$^8$ rad/s at T = 0 K. Displacements are measured in x-- and y-- directions. }}
\label{sfig2}
\end{figure}

\begin{figure*}
\includegraphics[width=1.0\textwidth,angle=0]{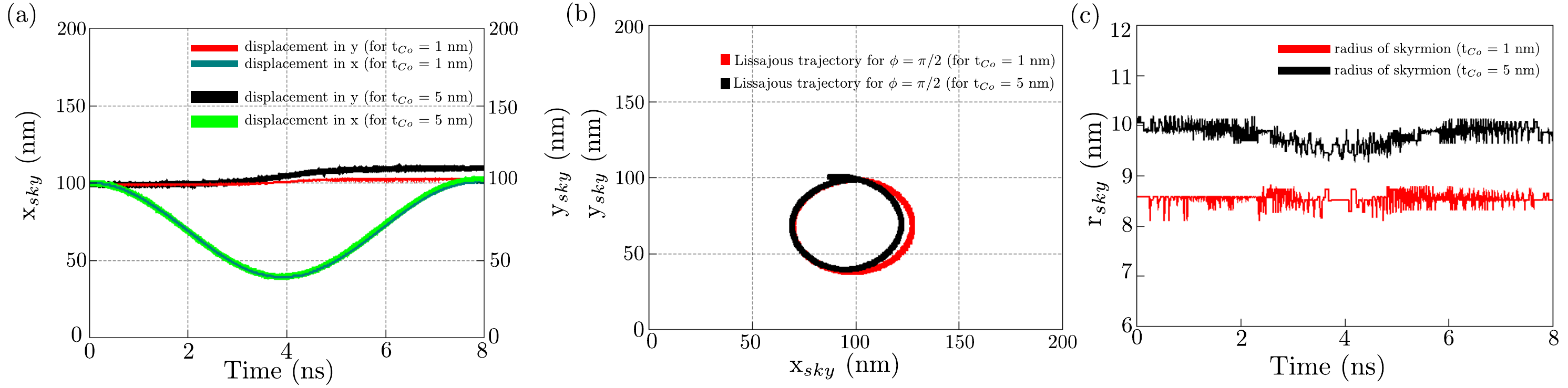}
\caption{{The skyrmions simulated in Co/Pt system for t$_{Co}$ = 1 nm and t$_{Co}$ = 5 nm are subjected to motion under the application of (a) $\bm{j}$ = (Asin($\omega t$), 0, 0), and (b)$\bm{j}$ = (Acos($\omega t$), Asin($\omega t$), 0). (c) The skyrmions' size variation during the entire simulation time from t = 0 to 8 ns.} }
\label{sfig3}
\end{figure*}

\subsection{Effect of size of the skyrmion on oscillation}
Keeping all the material parameters same, we change $D_{int}$ from 3 $\times$ 10$^{-3}$ J/m$^2$ to 3.4 $\times$ 10$^{-3}$ J/m$^2$. An ac pulse of $\bm{j}$ = (Asin($\omega$t), 0, 0) is applied to the sample, where A = 5 $\times$ 10$^{11}$ A/m$^2$ and $\omega$ = 8 $\times$ 10$^{8}$ rad/s. We demonstrate the result in fig. \ref{sfig2} where the skyrmion's radii are 8.5 nm (for $D_{int}$ = 3.0 $\times$ 10$^{-3}$ J/m$^2$) and 14.6 nm (for $D_{int}$ = 3.4 $\times$ 10$^{-3}$ J/m$^2$). The displacement of the skyrmion from the centre of the nano-structure is same in the direction of applied current, and A$_{sky}$ (x-- direction) is found to be 30.48 nm for both the cases. However, the bigger skyrmion shows a displacement in y-- direction too. A$_{sky}$ (y-- direction) for the skyrmion of radius = 14.6 nm is calculated to be 5.5 nm.

\begin{figure*}
\includegraphics[width=1.0\textwidth,angle=0]{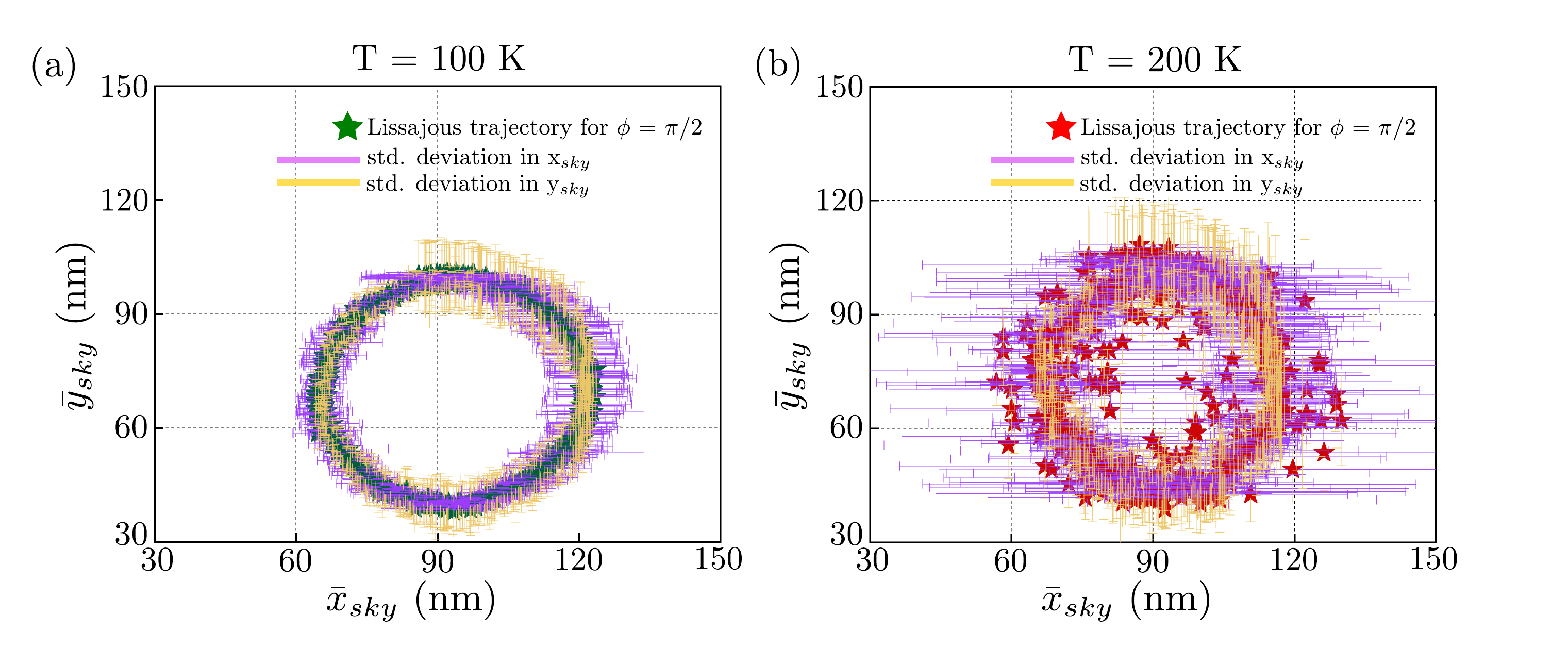}
\caption{{The mean Lissajous path of the skyrmion is presented for (a) T = 100 K and (b) T = 200 K with the error $\pm \sigma$. The applied current density is $\bm{j}$ = (Acos($\omega t$), Asin($\omega t$), 0) where, A = 5 $\times$ 10$^{11}$ A/m$^2$ and $\omega$ = 5 $\times$ 10$^{8}$ rad/s. The light purple and yellow  lines represent errors in in x$_{sky}$ and y$_{sky}$, respectively. From (a) to (b) as the temperature increases the error also increases.}}
\label{sfig4}
\end{figure*}

\subsection{Effect of free layer thickness on the skyrmion dynamics}

To understand the effect of free layer thickness(t$_{Co}$) on the skyrmion dynamics when subjected to an ac drive, we perform the simulation at t$_{Co}$ = 5 nm and obtain a skyrmion of radius $~$ 10 nm. We apply a current pulse of (a) $\bm{j}$ = (Asin($\omega t$), 0, 0), and (b) $\bm{j}$ = (Acos($\omega t$), Asin($\omega t$), 0). Fig. \ref{sfig3}a refers the oscillation of both the skyrmions simulated at t$_{Co}$ = 1 nm and 5 nm in x-- direction. We notice the x-- displacements are same for both the skyrmions. However, a slight displacement in y-- direction is observed for the skyrmion at t$_{Co}$ = 5 nm which is not present for the skyrmion nucleated at t$_{Co}$ = 1 nm. Fig. \ref{sfig3}b illustrates the skyrmion Lissajous trajectories for different free-layer thicknesses, confirming that the skyrmion continues to exhibit Lissajous motion across the entire range of thicknesses investigated.  Fig. \ref{sfig3}c outlines the change in radius of the skyrmion from t$_{Co}$ = 1nm to t$_{Co}$ = 5 nm.

\begin{figure*}
\includegraphics[width=1.0\textwidth,angle=0]{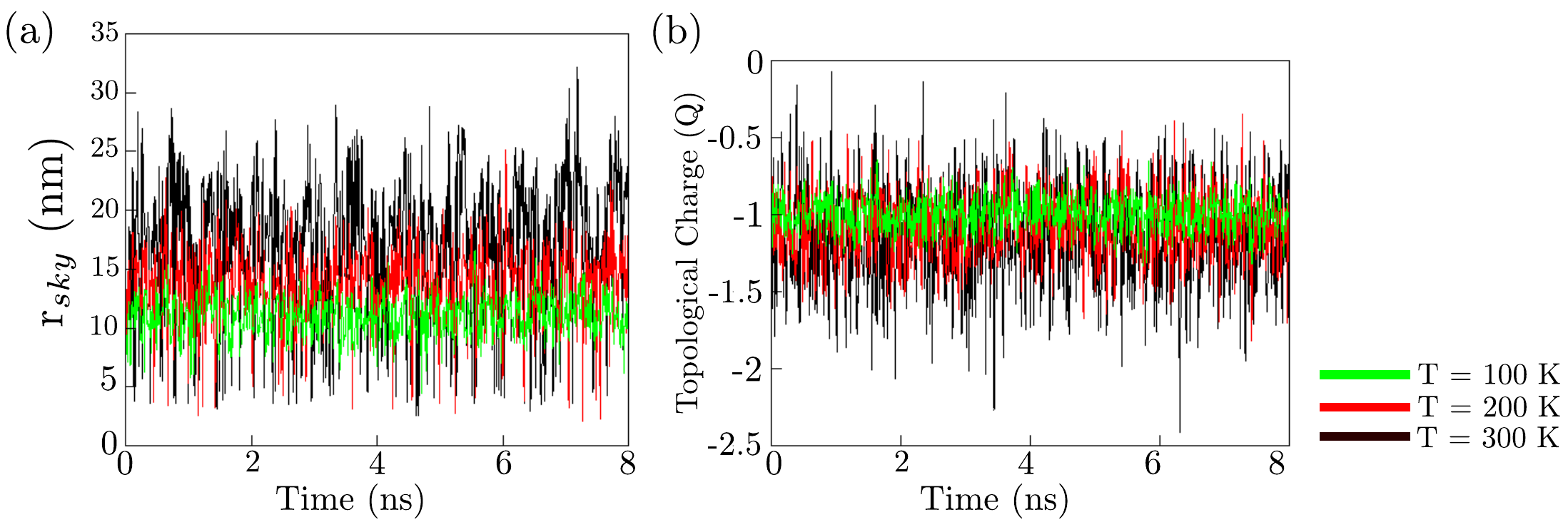}
\caption{{The variation of (a) radius (r$_{sky}$) and (b) topological charge of the skyrmions simulated at T = 100 K, 200 K and 300 K is presented with green, red and black lines respectively. Increasing the temperature from T = 100 K to 300 K leads to larger fluctuations in both r$_{sky}$ and Q.} }
\label{sfig5}
\end{figure*}


\end{document}